\documentclass[
aps,%
11pt,%
final,%
notitlepage,%
oneside,%
twocolumn,%
nobibnotes,%
nofootinbib,%
superscriptaddress,%
noshowpacs,%
centertags]%
{revtex4}
\usepackage{longtable}

\begin{document}
%
%
%



\selectlanguage{english}
\title{Catalog of Isolated Galaxies Selected from the 2MASS
Survey}

\author{\firstname{V.~E.}~\surname{Karachentseva}}
\affiliation{Main Astronomical Observatory National Academy of Sciences, Akad.
Zabolotnogo, 27, Kiev, Ukraine}
\author{\firstname{S.~N.}~\surname{Mitronova}}
\affiliation{\saoname}
\author{\firstname{O.~V.}~\surname{Melnyk}}
\affiliation{Institut d'Astrophysique et de G\'eophysique,
Universit\'{e} de Li\`{e}ge, All\'{e}e du 6 Ao\^{u}t, 17, B5C,
Belgium} \affiliation{Astronomical Observatory, Taras Shevchenko
Kiev National University, Observatornaya 3, 04053, Ukraine}

\author{\firstname{I.~D.}~\surname{Karachentsev}}
\affiliation{\saoname}

\received{June 30, 2009}%
\revised{September 3, 2009}%

\begin{abstract}
We search for isolated galaxies based on the automatic
identification of isolated sources from the Two Micron All-Sky
Survey (2MASS) followed by a visual inspection of their
surroundings. We use the modified Karachentseva criterion to
compile a catalog of  3227 isolated galaxies (2MIG), which
contains 6\% of 2MASS Extended Sources Catalog (or 2MASX) sources
brighter than $K_{\rm s }=12^m$ with angular diameters $a_K \geq
30\arcsec$. The catalog covers the entire sky and has an effective
depth of $z\sim$ 0.02. The 2493 very isolated objects of the
catalog, which we include into the 2MVIG catalog, can be used as a
reference sample to investigate the effects of the environment on
the structure and evolution of galaxies located in regions with
extremely low density of matter.
\end{abstract}
 \maketitle

\section{INTRODUCTION}

Current observational data about the distribution of galaxies
demonstrate the existence of large-scale cosmic structures
consisting of filaments, walls, and clusters bordering empty
volumes of space (voids). Such a cellular structure can be
naturally explained in terms of the standard  $\Lambda$CDM
cosmological model. About 5--10\% of all galaxies reside in dense
(virialized) cluster regions, and about the same number is found
in the nonvirialized neighborhoods of these clusters. Most of the
galaxies are located in sparsely populated isolated systems
resembling the Local Group (50\%) or in low-density clouds (25\%)
of which the closest to us is Canes Venatici I. The remaining
5--10\% of galaxies are scattered in the general metagalactic
field. Note that the estimate of the fraction of  ``truly
isolated'' galaxies still remains a topic of debate.

Isolated galaxies are objects that have not been appreciably
affected by their closest neighbors over the past 1--2~Gyr. This
means that their physical properties are determined mostly by the
initial formation conditions and internal evolutionary processes.
A representative sample of isolated galaxies is needed to test
models of the formation and evolution of galaxies, and as a
reference sample for the studies of the properties of galaxies in
pairs, groups, and clusters in order to understand the effects of
the environment on such fundamental galactic properties as
morphology, gas and dust content, chemical composition, and
star-formation rate.

No catalog of isolated galaxies can be perfect. Our catalog, which is a magnitude- and angular-diameter
limited sample of galaxies, does not include dwarf objects in distant volumes, and some spatially
isolated galaxies may have close projected neighbors. On the other hand, such samples usually
contain a certain fraction of false isolated galaxies. When compiling a sample of isolated galaxies every
effort should be made to minimize these statistical errors. At the same time, the sample should be
representative, i.e., it must cover a significant fraction of the sky and be sufficiently deep.

The Catalog of Isolated Galaxies (known in Russian by its initials
\mbox{KIG) \cite{Kara:Mitronova_n}} is an example of a successful
attempt to compile such a sample. It is based on the following
empirical criteria used to select isolated objects:
\begin{equation}
         X_{1i}/a_i\geq s=20,
\end{equation}
\begin{equation}
       4 \geq a_i/a_1 \geq 1/4,
\end{equation}
where subscripts ``1'' and   ``{\it i}'' refer to the fixed galaxy
and its neighbors, respectively. According to these criteria, a
galaxy with a standard angular diameter  $a_1$  is considered
isolated if its angular separation $X_{1i}$ from all its neighbors
with ``significant'' angular diameters  $a_i$ inside interval (2)
is equal to or exceeds  $20 a_i$.  Criteria \mbox{(1)--(2)} were
applied \linebreak to 27 840 galaxies of the CGCG
catalog~\cite{Zwicky:Mitronova_n}, which contains northern-sky
galaxies with photographic magnitudes \mbox{$m_p < 15.7^m$}. As a
result of visual inspection of these galaxies and their
surroundings in DSS-1 images, a total of 1050 galaxies at Galactic
latitudes $\mid b\mid \geq 20\degr$, or 4\% of all CGCG galaxies
were found to meet the adopted isolation criterion.

The long-term AMIGA project \linebreak ({\tt
http://www.iaa.es/AMIGA.html/}) has been carried out since early
2000 by an international team from Spain, the United States,
France, Italy, and Germany. The aim of this project is to study
the physical properties of the most isolated galaxies of the KIG
and, especially, the properties of the interstellar medium in
these galaxies taking into account the observational data that
became available in recent years. In the course of these studies
the isolation criterion was tested and had its efficiency
confirmed for most of the KIG galaxies.

In recent years, attempts have been made to compile new catalogs of isolated galaxies based,
in particular, on the SDSS survey \cite{Allam:Mitronova_n}. However, so far these samples
cover only a small fraction of the sky.

In this paper we report a new catalog of isolated galaxies,
designated 2MIG, where we use the advantages offered by the
photometrically homogeneous 2MASS survey covering the entire
Northern and Southern hemispheres.

\section{SEARCH FOR ISOLATED GALAXIES IN THE 2MASS SURVEY}

The Two Micron All-Sky Survey (2MASS) \cite{Skrutskie:Mitronova_n}
was made in three infrared photometric bands, $J$
(1.11--1.36\,$\mu$m), $H$ (1.50--1.80\,$\mu$m), and $K_{\rm s}$
\linebreak \mbox{(2.00--2.32\,$\mu$m)}. It contains about  2.6
million extended sources with $K_{\rm s}$-band magnitudes brighter
than 14.5$^m$. As a result of this survey a total of 1.64 million
galaxies with $K_{\rm s } \leq 14.5^m$ and angular diameters
greater than 10$\arcsec$ have been identified, which now
constitute the  2MASS Extended Sources Catalog (XSC)
\cite{Jarrett:Mitronova_n}. For the 2MASS XSC objects, a large
number of structural and photometric parameters are listed, which
have been determined using homogeneous procedures, and that
explains why 2MASS XSC was used to create a number  of new
catalogs, e.g., the 2MFGC catalog of flat galaxies
\cite{Mit01:Mitronova_n}.

We identified isolated galaxies by applying slightly modified
criteria (1) and (2) to  the objects of the XSC 2MASS catalog. We
increased to $s=30$ the dimensionless  ``distance''  $X_{1i} /a_i
= X_{1i} /2r_i$ to the neighboring object in criterion (1), as the
infrared diameters of galaxies in the 2MASS are systematically
smaller than their standard optical
diameters~\cite{Jarrett01:Mitronova_n}. Thus we found that the
median standard blue to infrared diameter ratio for  RFGC
\cite{Karach03:Mitronova_n} galaxies was $a_{25} /2r_{20fe} = 1.5$
and it differed widely for galaxies of different \mbox{structures
\cite{Karach04:Mitronova_n}}. For example, we consider galaxy
``1'' with a  $K$-band magnitude \mbox{$K_{20fe} \equiv K_{\rm
s}$} and isophotal  $K$-diameter $a_K\equiv 2r_{20fe}$ to be
isolated if conditions (1) and (2) are fulfilled for this galaxy
and any of its significant neighbors for $s=30$. When testing the
isolation of a galaxy with respect to its possible faint
companions, we apply the algorithm of isolated galaxy
identification to all candidate galaxies with apparent $K_{\rm
s}$-band magnitudes in the interval
\begin{equation}
      4.0^m< K_{\rm s} \leq 12.0^m
\end{equation}

\noindent
and angular diameters
\begin{equation}
a_K \geq 30\arcsec.
\end{equation}

We chose the limiting apparent magnitude \linebreak \mbox{$K_{\rm
s }=12.0^m$} so as to make it consistent with the limiting
magnitude of the KIG for galaxies with typical color indices
$B-K_{\rm s}= 3.5^m - 4.0^m$. Note that below the \mbox{$K_{\rm s
}=12.0^m$} limit and down to the 2MASS limiting magnitude there
remain many faint galaxies with \mbox{$K_{\rm s }=12.0^m-14.5^m$,}
which we use in the test for isolation. The bright-end magnitude
constraint is due to the specifics of the 2MASS photometry of the
most extended bright \mbox{galaxies \cite{Jarrett:Mitronova_n}.}
The 2MASS XSC catalog does not include objects with angular
diameters \mbox{$a_K= 2 r_{20fe} <10\arcsec$}, and therefore for
the condition (2) to be fulfilled near its lower limit, we also
had to limit the angular diameters of potential isolated galaxy
candidates. We set the minimum diameter equal to 30$\arcsec$, so
that our sample would remain sufficiently representative, although
conditions (2) and (4) become somewhat inconsistent for galaxies
with small diameters \mbox{$a_K=30\arcsec$--$40\arcsec$}.

We performed an automatic identification of isolated galaxies
using the Pleinpot software package \linebreak ({\tt
http://leda.univ-lyon1.fr/pleinpot/\linebreak /pleinpot.html})
developed for the reduction and analysis of astronomical data. For
each object and its nearest neighbor from the 2MASX we fix the
coordinates, $K_{\rm s}$-band magnitudes, and  $a_K/2$ radii. All
neighbors with significant sizes located in the
$31\degr\times31\degr$ area centered on each of the 51572
candidate isolated galaxies were identified, and the mutual
distances between the fixed and neighboring galaxies were found.
Significant neighbors of the galaxy considered were then ranked by
the dimensionless distance $s$. For a galaxy to be classified as
isolated its dimensionless distance from all the significant
neighbors must exceed \mbox{$s=30$.} We chose the above areas with
a 1$\degr$ overlap and identified and excluded all the shared
galaxies found at the boundaries between the adjacent areas. An
automatic algorithm produced a sample of 4045 objects, which we
then subjected to a visual inspection.

Near-infrared photometry is superior to optical surveys in that it
is affected only slightly by the galactic extinction. The major
disadvantage of the 2MASS survey is short exposure time (about
7\,s/object), which made the survey insensitive to low surface
brightness objects, especially to those of blue color. As a
result, many dwarf irregulars are unrepresented in the 2MASS
survey and this may seriously affect the decision whether a
particular galaxy is isolated or not. That is why at the second
stage of our work we visually inspected the neighborhood of each
of the 4045 selected galaxies on the DSS-1 and DSS-2 images,
paying special attention to those objects that are absent from the
2MASS.
\begin{figure*}[tbp]
\setcaptionmargin{5mm}
\onelinecaptionstrue
\includegraphics[width=15cm, bb=23 481 485 816, clip]{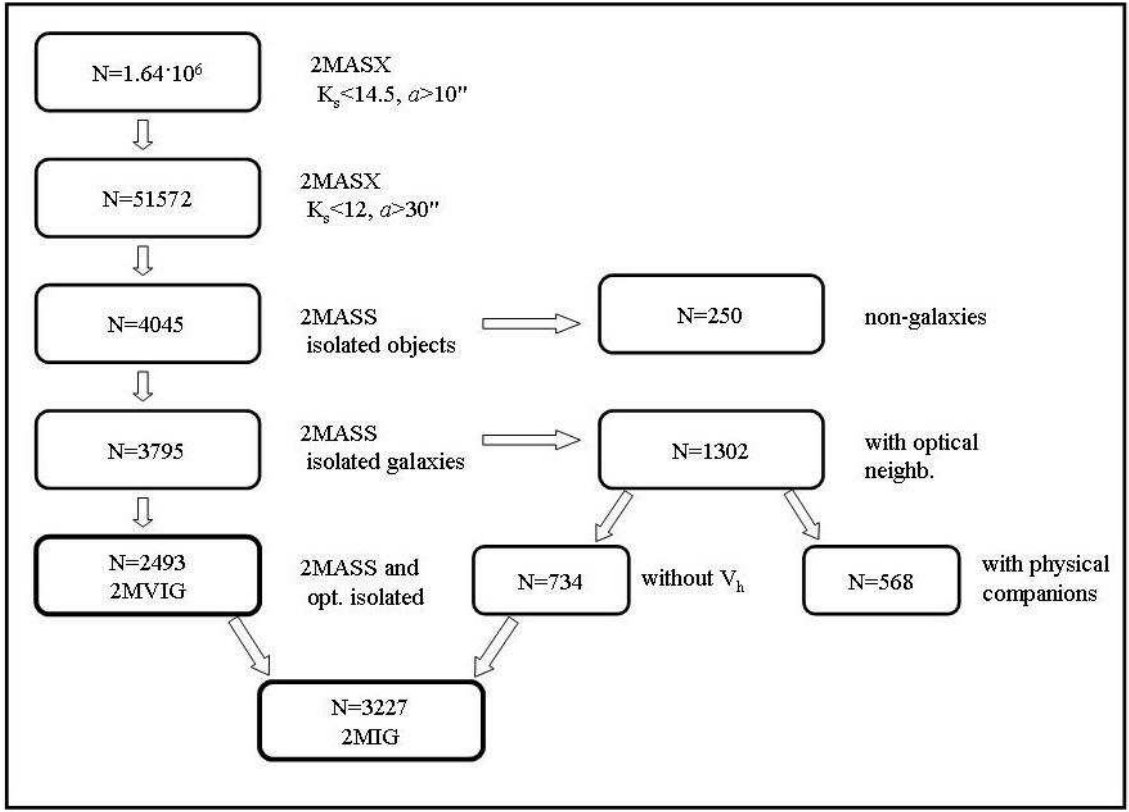}
\bigskip
\captionstyle{normal}
\caption{Selection of isolated galaxies.}
\end{figure*}

At first we identified among the 4045 extended objects those that
we found not to be galaxies, but rather planetary nebulae, star
clusters, or multiple stars ($N = 250$). We excluded these objects
from further analysis. We then selected in DSS-1 a
\mbox{$20\arcmin\times20\arcmin$} to $60\arcmin\times60\arcmin$
wide area (depending on the diameter $a_1$) around each of the
remaining 3795 galaxies, measured the optical angular diameters of
the neighboring galaxies in this area, and marked all the
significant members according to Karachentseva's criterion (1) and
(2). We considered a galaxy to be  ``very isolated'' if the
inspection of its neighborhood revealed no significant neighbors
either in the infrared or at optical wavelengths. As a result, we
found a total of 2493 such ``2MASS Very Isolated Galaxies''
(2MVIG), which make up a total of 4.8\% of all galaxies brighter
than \mbox{$K_{\rm s }=12.0^m$} with diameters $a_K \geq
30\arcsec$. If we found the candidate galaxy to have significant
neighbors in optical images, we additionally checked their radial
velocities using the LEDA and NED databases and the catalogs of
the pairs and groups of \mbox{galaxies
\cite{KarMak:Mitronova_n,MakKar:Mitronova_n}.} A total of 568 of
the 1302 galaxies with one or several significant neighbors found
on DSS images have neighbors with radial velocities close to that
of the candidate galaxy ($\Delta V_h <500$ km/s). No radial
velocity measurements are available for the significant projected
members of the remaining 734 candidate isolated galaxies. A
considerable part of these objects may eventually make it into the
list of truly isolated galaxies. We came to this conclusion after
comparing the radial velocities of the galaxy and of its nearest
neighbor and after analyzing the results of our radial velocity
measurements of the neighbors of isolated galaxies in the Local
Supercluster \cite{Melnik:Mitronova_n}. We refer to the combined
sample of these 734 galaxies and 2493 2MVIG galaxies as the  2MASS
Isolated Galaxies Catalog, or 2MIG. Figure~1 shows schematically
the procedure of creating the 2MIG and 2MVIG samples.
Figures~2a--2e present examples of isolated galaxies identified in
the 2MASS and located in different environments. The images are
extracted from the DSS-1 survey. The areas have a size of
$20\arcmin\times20\arcmin$: North is at the top and East on the
left. Candidate \mbox{isolated} galaxies are located at the
center, and their neighbors are marked by the arrows. Let us now
give brief comments on various situations found as a result of the
optical inspection of the candidate isolated galaxies. Figure\,2a:
the galaxy has no significant neighbors neither in the 2MASS nor
in the DSS survey. In Fig.~2b: the neighboring galaxy does not
show up in the 2MASS, but is significant in the DSS. According to
the quoted radial velocities, it is a foreground galaxy with no
effect on the isolation of the candidate galaxy\footnote{The 2MIG
938 galaxy belongs to this group. Formally its nearest neighbor
from the 2MASX list is located at a distance of  $2s=65$. However,
visual inspection reveals another galaxy---KPG 123A---to be in
contact with  KPG 123B. KPG 123A has a similar optical size and
has been detected as a low surface brightness infrared source in
2MASXi. Given that the radial velocities of these galaxies differ
by  900 km/s or 700 km/s according to the NED and LEDA,
respectively, we can consider KPG 123B to be an isolated galaxy.}.
Figure~2c: the neighboring galaxy cannot be seen in the 2MASS, but
is significant in the DSS. We exclude the tested galaxy from the
list of isolated galaxies because the two objects have close
radial velocities. Figure~2d: the neighboring galaxies that are
significant in optical images have no radial velocity estimates
available. In these cases we considered the candidate galaxy as
possibly isolated.

Lastly we should mention a special case, where a 2MASX galaxy has
no significant members, but has dwarf objects found in its
vicinity, which, albeit ``insignificant'' according to the
criterion (1)--(2), have radial velocities close to that of the
galaxy considered. Figure~2e shows an example of such a system
(note that the field contains other bluish objects with no radial
velocity measurements). We formally consider such galaxies to be
isolated. Systems consisting of a solitary galaxy surrounded by
dwarf companions exclusively are by themselves of considerable
interest from the viewpoint of their occurrence and dynamical
evolution. ``Insignificant'' companions with close radial
velocities have been found for  5\% of the galaxies included into
the 2MVIG catalog.

\begin{figure*}[tbp]
\setcaptionmargin{5mm}
\onelinecaptionsfalse
\includegraphics[width=7cm]{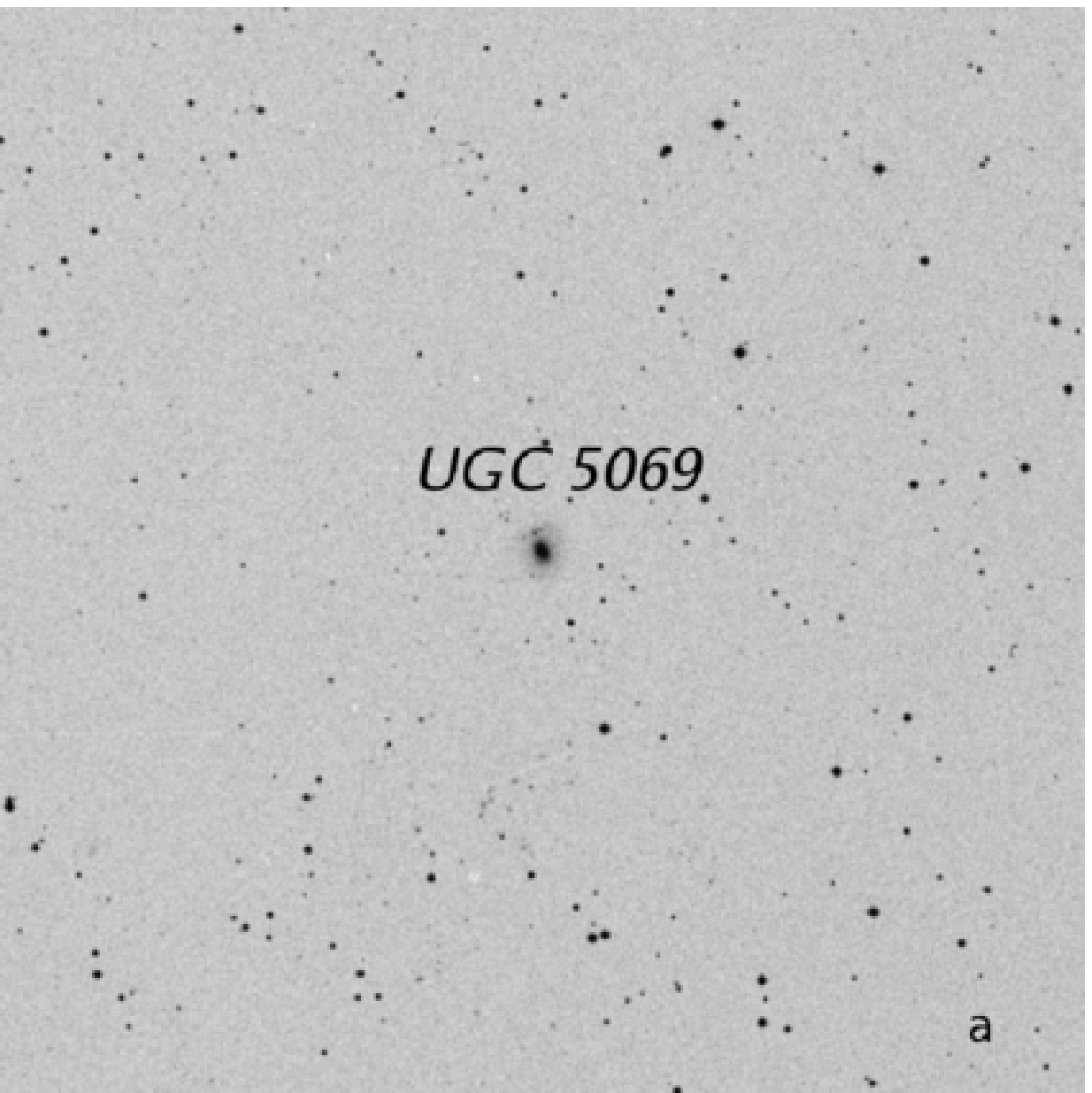}
\includegraphics[width=7cm]{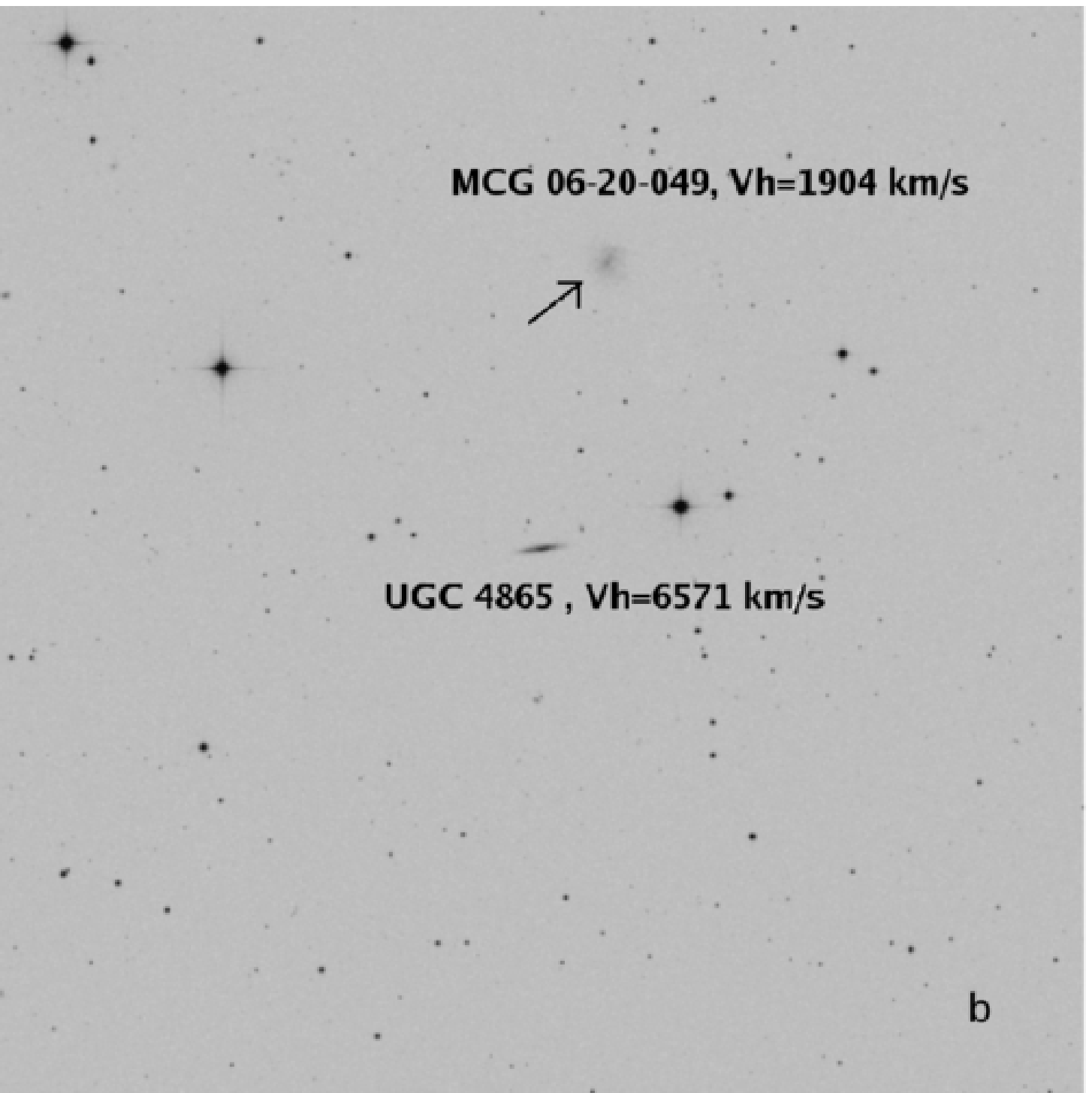}
\includegraphics[width=7cm]{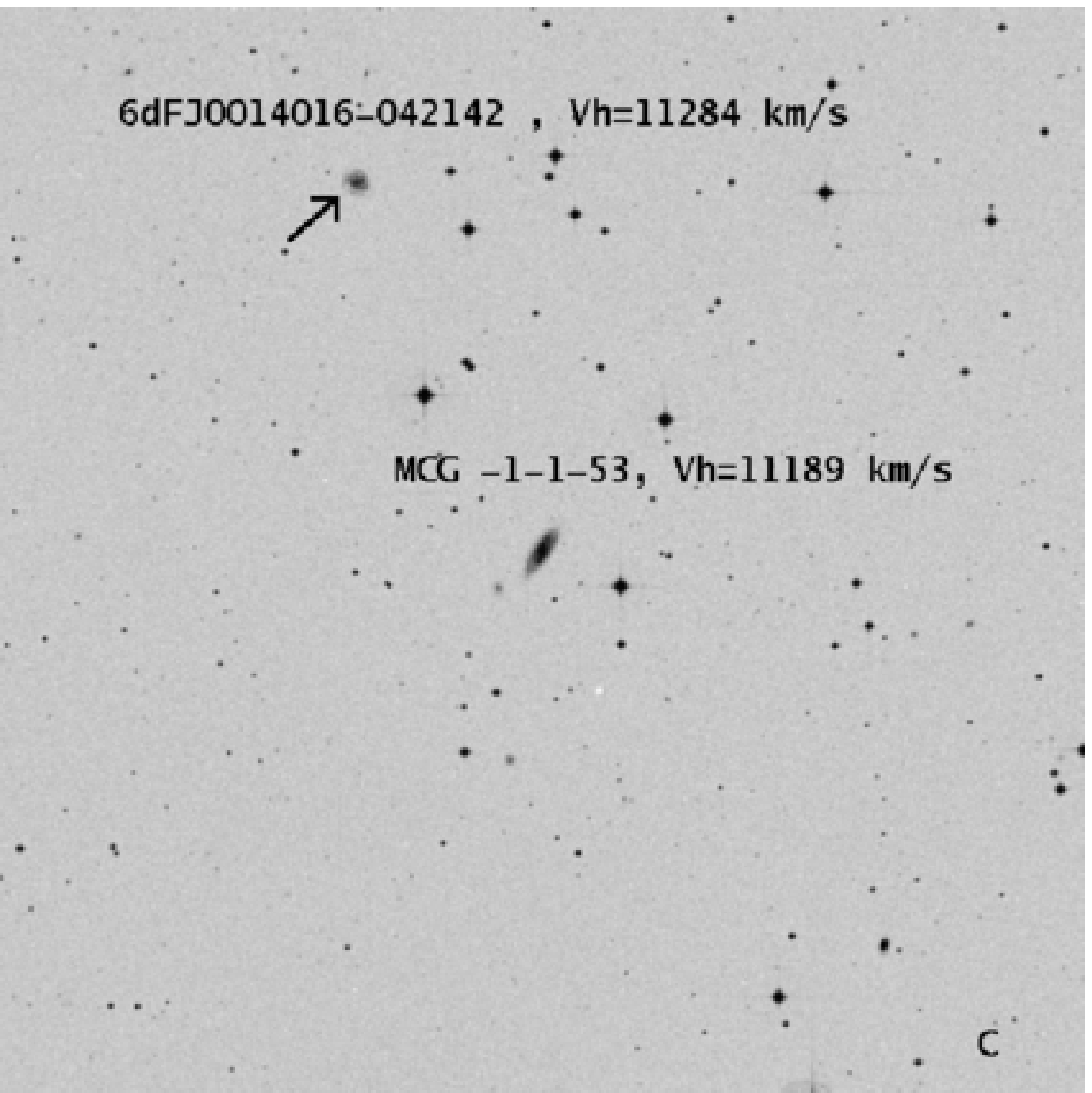}
\includegraphics[width=7cm]{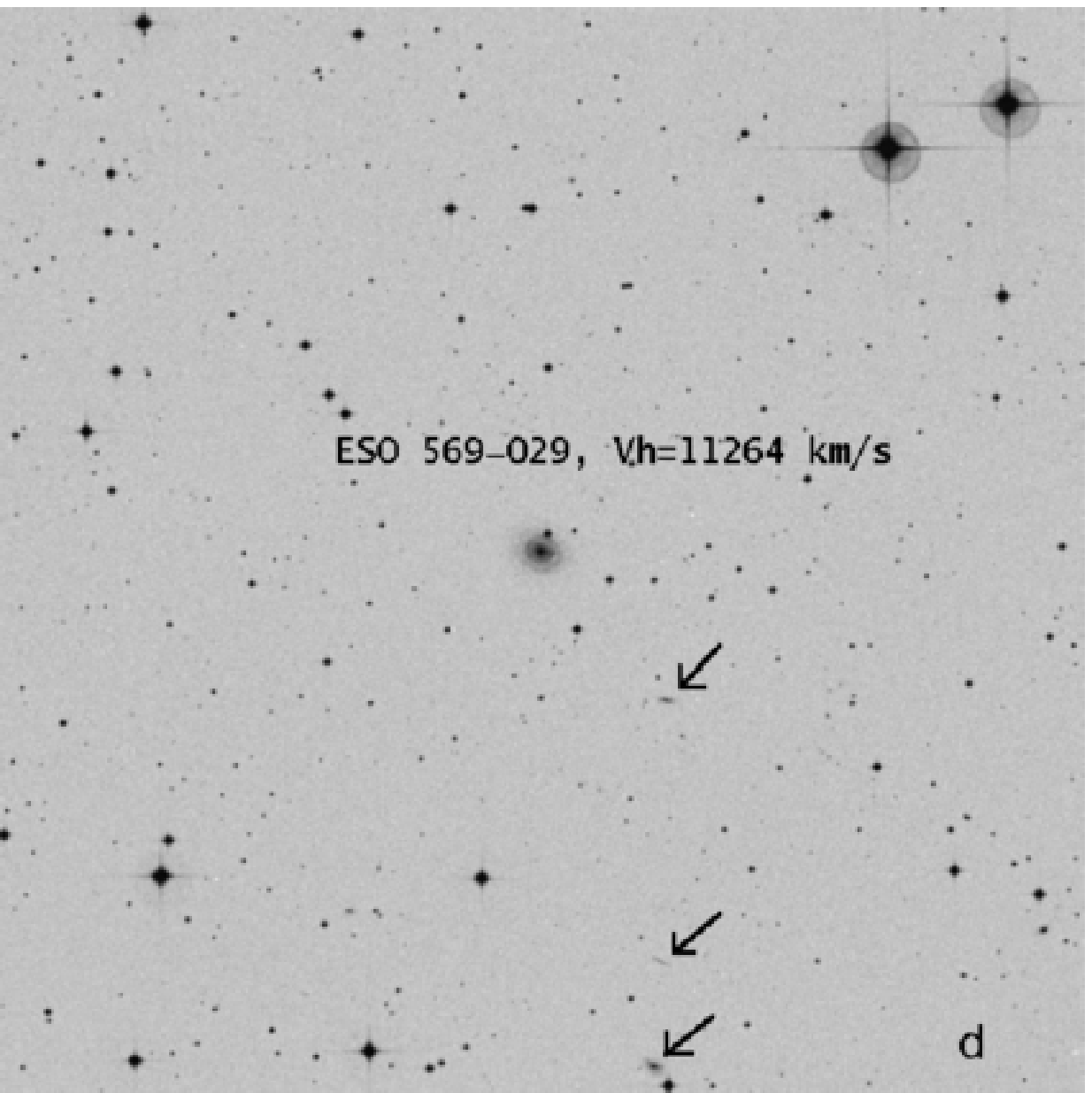}
\includegraphics[width=7cm]{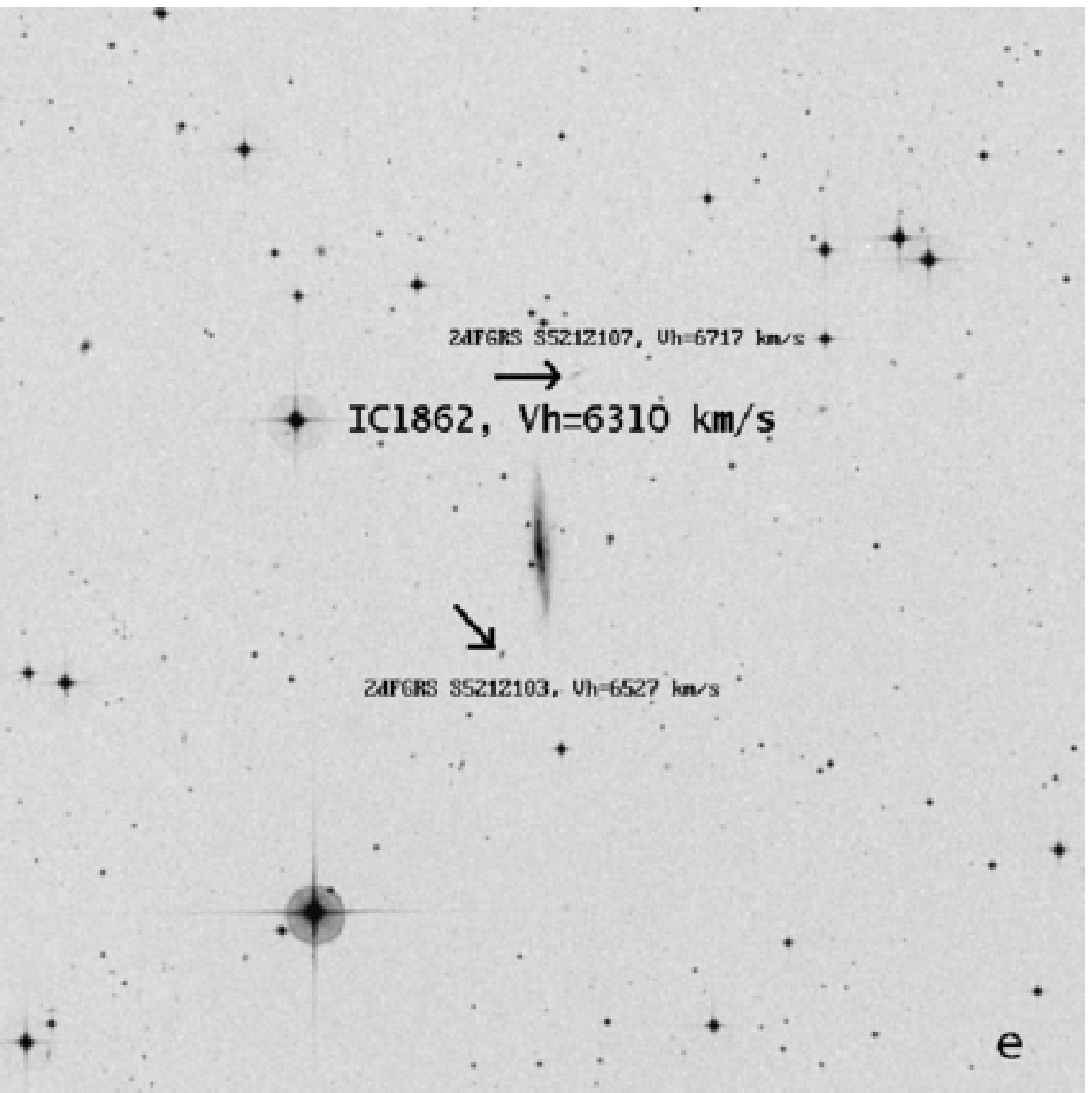}
\captionstyle{normal}
\caption{Examples of isolated galaxies in different environments. The arrows indicate the neighboring
galaxies.}
\end{figure*}

\section{2MIG CATALOG OF ISOLATED GALAXIES}

Based on the above considerations, we make our combined 2MIG
catalog of isolated galaxies (which is partly listed in the Table)
of a total of  3227 objects including the 2MVIG sample of 2493
among most isolated galaxies. The electronic version of the
catalog is available at {\tt
ftp://cdsarc.u-strasbg.fr/pub/\linebreak /cats/VII/257}.

\begin{figure*}[tbp]
\setcaptionmargin{5mm}
\onelinecaptionstrue
\includegraphics[width=13cm, bb=11 13 282 219, clip]{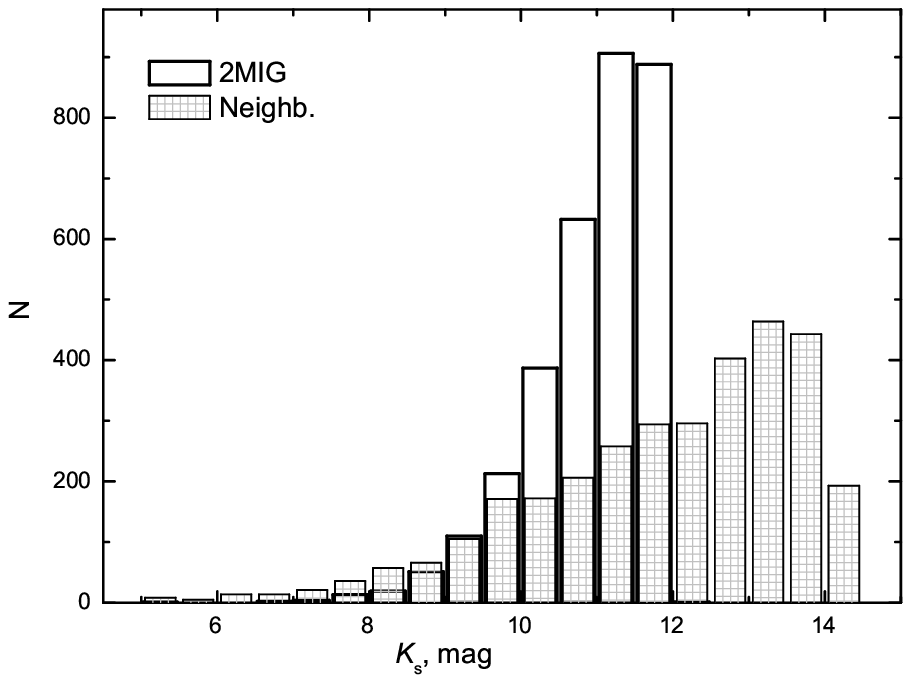}
\captionstyle{normal}
\caption{Distribution of $K_{\rm s}$-band magnitudes of 2MIG isolated galaxies and their closest
neighbors.}
\end{figure*}

The columns of the Table give:

(1) the running number of the object in the 2MIG catalog;

(2) the 2MASS equatorial coordinates for J2000.0 epoch;

(3) name of the galaxy according to the NED database. If the
galaxy could not be identified in the known catalogs, we list it
in column (3) as a 2MASX source with the corresponding coordinates
in column~(2) or under a PGC number from the LEDA database;

(4) $r_{20fe}$ is angular radius (semimajor axis) in arcsec;

(5) $K_{\rm s}$ is the 2MASS magnitude corresponding to the
$K_{20fe}$ isophote;

(6)  $2s=  X_{1i}/r_i$ is a dimensionless  ``distance'' between
the isolated galaxy and its nearest significant neighbor;

(7) heliocentric radial velocity  (in km/s) adopted from the LEDA
or NED database. A total of  2328 (72\%) and 1775 (71\%) galaxies
in the 2MIG and 2MVIG catalogs, respectively, have measured radial
velocities;

(8) morphological type encoded in the de Vaucouleurs scale. We
visually estimated the morphological types of the catalog galaxies
by inspecting their images in the DSS-1, DSS-2, and SDSS optical
surveys and the 2MASS $JHK$-band images of the central parts of
the galaxies. Morphological types could be determined with less
certainty in the zone of strong extinction at low Galactic
latitudes \mbox{$\mid b\mid<10\degr$}. We adopted the same $T =
-2$ for the elliptical galaxies of all subclasses irrespective of
their apparent ellipticity;

(9) the number of significant neighbors of the isolated galaxy
found as a result of further inspection of its vicinity on the
DSS-1 images; the blanks correspond to 2MVIG galaxies with no
significant neighbors found on optical images;

(10) comments on the morphological peculiarities of an isolated
galaxy and its identification in the NED with the objects from the
KIG, IRAS, and catalogs of active and peculiar galaxies (by
Markarian, Arp and Madore, etc). The comments here as well
indicate the presence of insignificant companions with close
($\Delta V_h \leq 500$ km/s) radial velocities. Vc indicates a
companion with velocity.

\begin{figure*}[tbp]
\setcaptionmargin{5mm}
\onelinecaptionsfalse
\includegraphics[width=13cm, bb=12 12 288 220, clip]{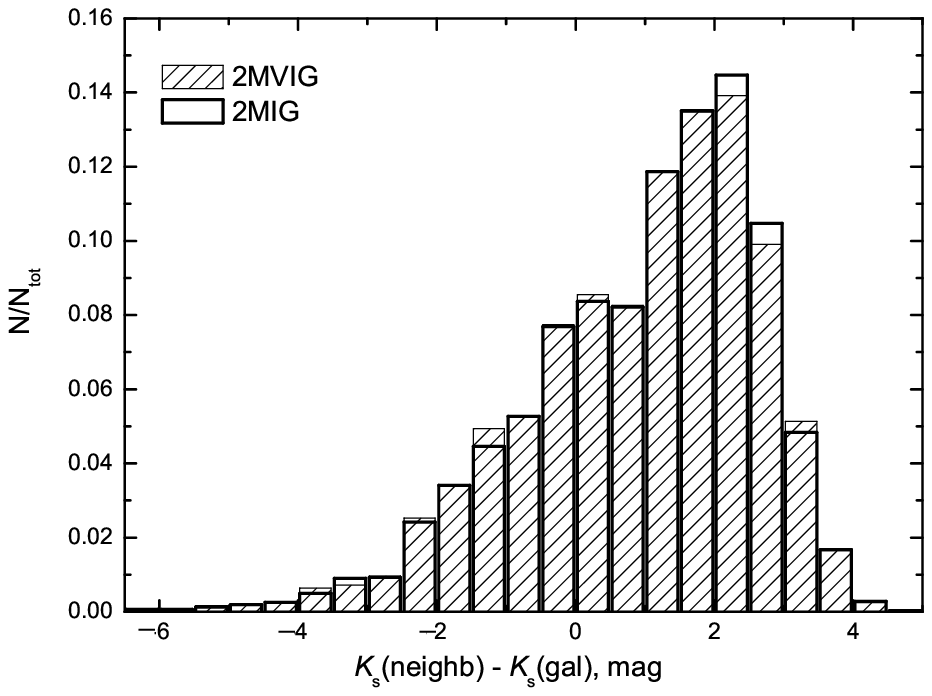}
\captionstyle{normal} \caption{ Distribution of the ``nearest
significant neighbor vs. isolated galaxy'' magnitude differences
for 2MIG and 2MVIG galaxies.}
\end{figure*}

We identified a total of 244 KIG galaxies, 1053 infrared IRAS
sources, and 94 galaxies from the catalogs and lists of active
galaxies among the 2MIG objects. Among the 2MVIG objects, we found
a total of 227, 820, and 69 KIG galaxies, IRAS sources and active
galaxies, respectively. A significant fraction of isolated
galaxies exhibits such features as a bar, ring, or an asymmetric
(peculiar) structure.

\section{2MIG GALAXIES AND THEIR SIGNIFICANT NEIGHBORS}

Let us now compare some parameters of isolated 2MIG galaxies and
of their nearest significant neighbors found in the 2MASS XSC
survey.

\begin{figure*}[tbp]
\setcaptionmargin{5mm}
\onelinecaptionsfalse
\includegraphics[width=13cm, bb=10 10 291 218, clip]{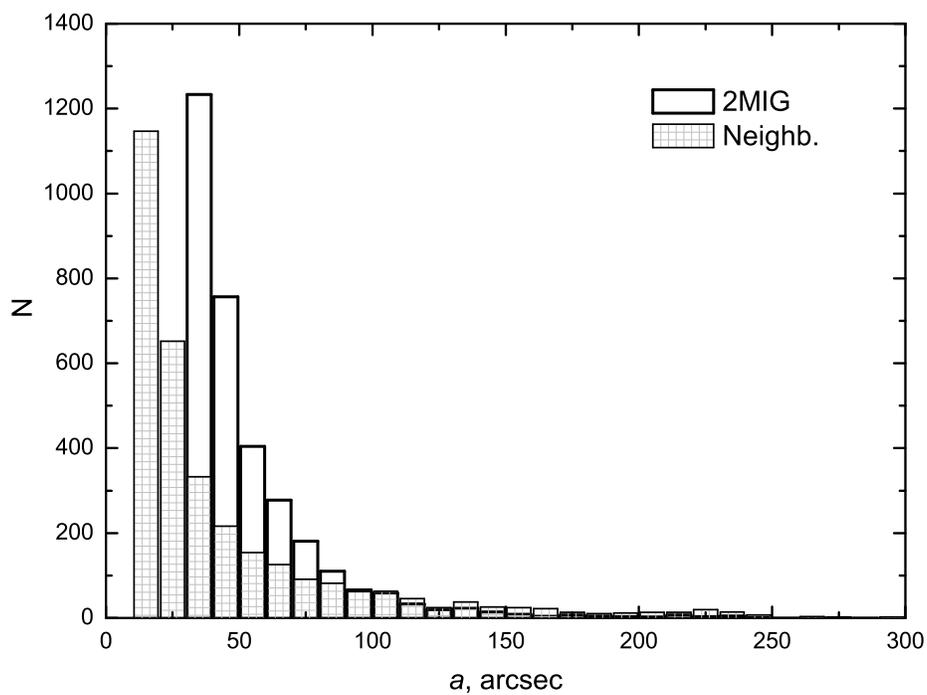}
\captionstyle{normal} \caption{Angular diameter distributions of
the 2MIG galaxies and their nearest significant neighbors.}
\end{figure*}

Figure~3 presents differential distributions of $K_{\rm s}$
magnitudes of the 2MIG isolated galaxies and their nearest
significant neighbors. The mean apparent magnitude of the sample
of isolated galaxies and its standard deviation are \mbox{$\langle
K_{\rm s}\rangle =10.94^m$} and  \mbox{$SD=0.81^m$,} respectively.
The corresponding parameters for the 2MVIG sample are almost the
same and equal to $10.90^m$ and $0.84^m$, respectively. Magnitude
distributions of the nearest significant neighbors have an
appreciably greater scatter and are shifted towards fainter
objects:  $\langle K_{\rm s}\rangle=11.92^m$ and $SD=1.76^m$.

\begin{figure*}[tbp]
\setcaptionmargin{5mm}
\onelinecaptionsfalse
\includegraphics[width=13.3cm, bb=22 22 308 234, clip]{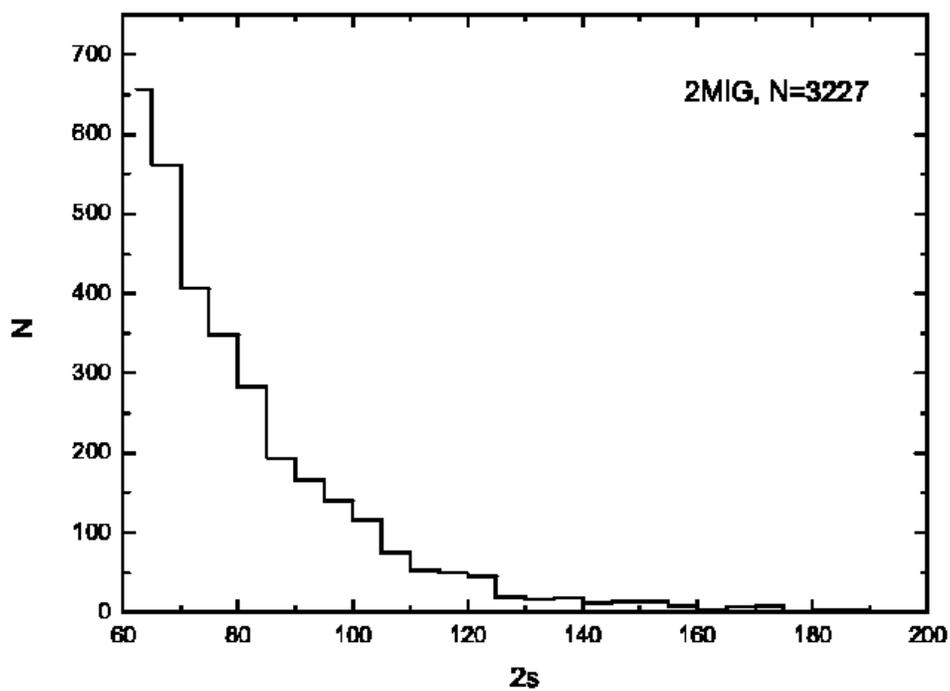}
\captionstyle{normal}
\caption{Distribution of dimensionless parameter $2s$ for the nearest neighbors of  3227  2MIG galaxies.}
\end{figure*}

\begin{figure*}[tbp]
\setcaptionmargin{5mm}
\onelinecaptionsfalse
\includegraphics[width=13cm, bb=11 12 285 217, clip]{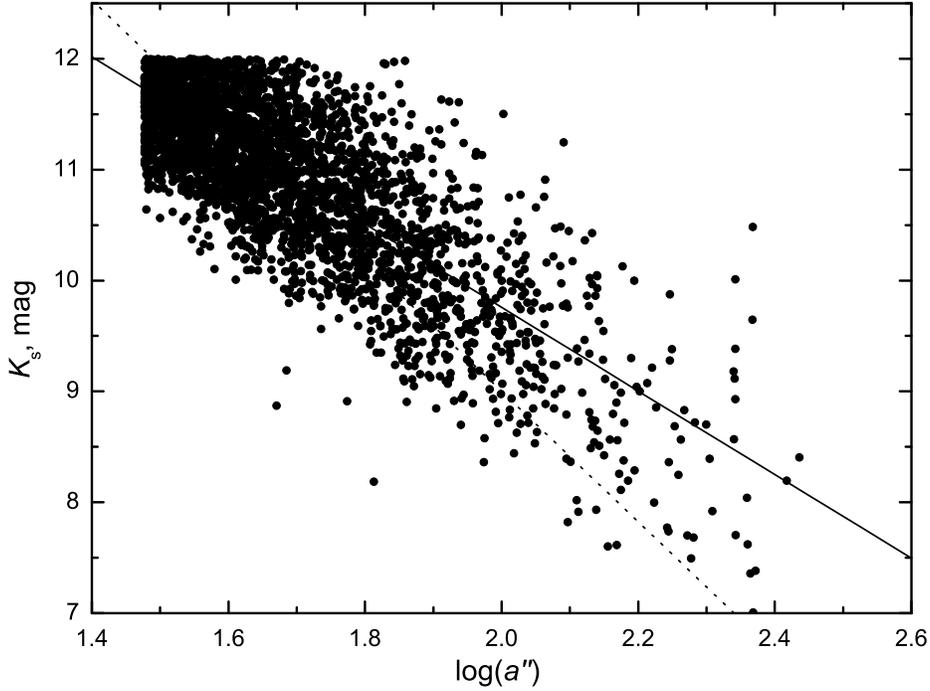}
\captionstyle{normal} \caption{Relation between the apparent
$K$-band magnitude and logarithm of angular diameter for 3227
galaxies of the 2MIG catalog. The solid and dashed lines show the
``$K_{\rm s}$ vs. $a$'' and ``$a$ vs. $K_{\rm s}$'' regressions,
respectively.}
\end{figure*}

\begin{figure*}[tbp]
\setcaptionmargin{5mm}
\onelinecaptionsfalse
\includegraphics[width=12cm, bb=7 11 267 223, clip]{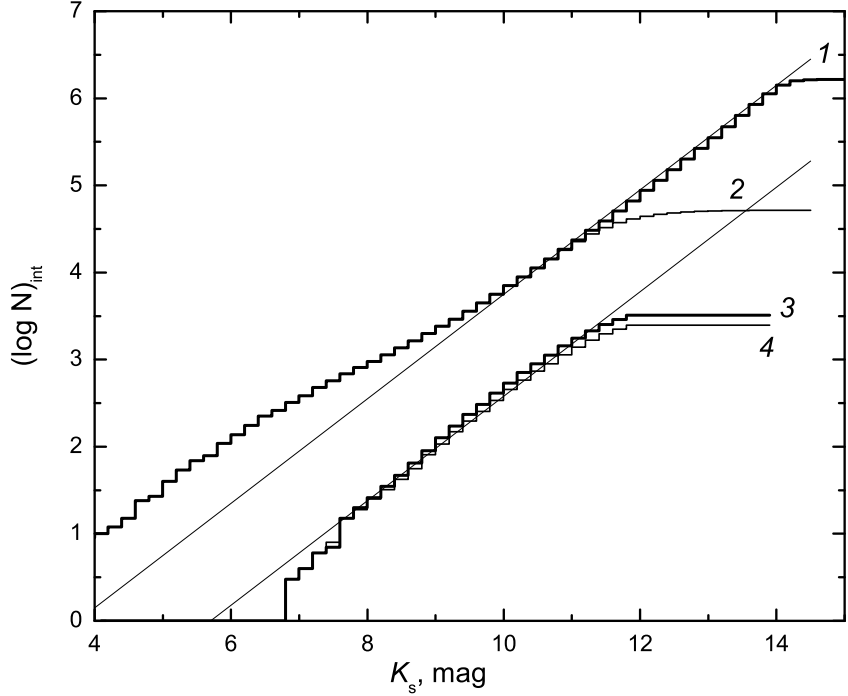}
\captionstyle{normal} \caption{The ``number of galaxies vs.
apparent magnitude'' relation  for all 2MASS XSC galaxies ($N =
1.6\times 10^6$) ({\it 1}); for the 2MASS  XSC  sample with
$K_{\rm s }\leq12^m, a_K \geq 30\arcsec$ ($N = 51572$) ({\it 2});
for the 2MIG catalog of isolated galaxies ($N = 3227$) ({\it 3}),
and  for the 2MVIG sample of  ``truly'' or ``very isolated''
galaxies ($N = 2493$)  ({\it 4}). The straight lines correspond to
the uniform distribution  $\log N( < K_{\rm s}) \propto 0.6K$.}
\end{figure*}

\begin{figure*}[tbp]
\setcaptionmargin{5mm}
\onelinecaptionsfalse
\includegraphics[width=16cm, bb=20 538 508 793, clip]{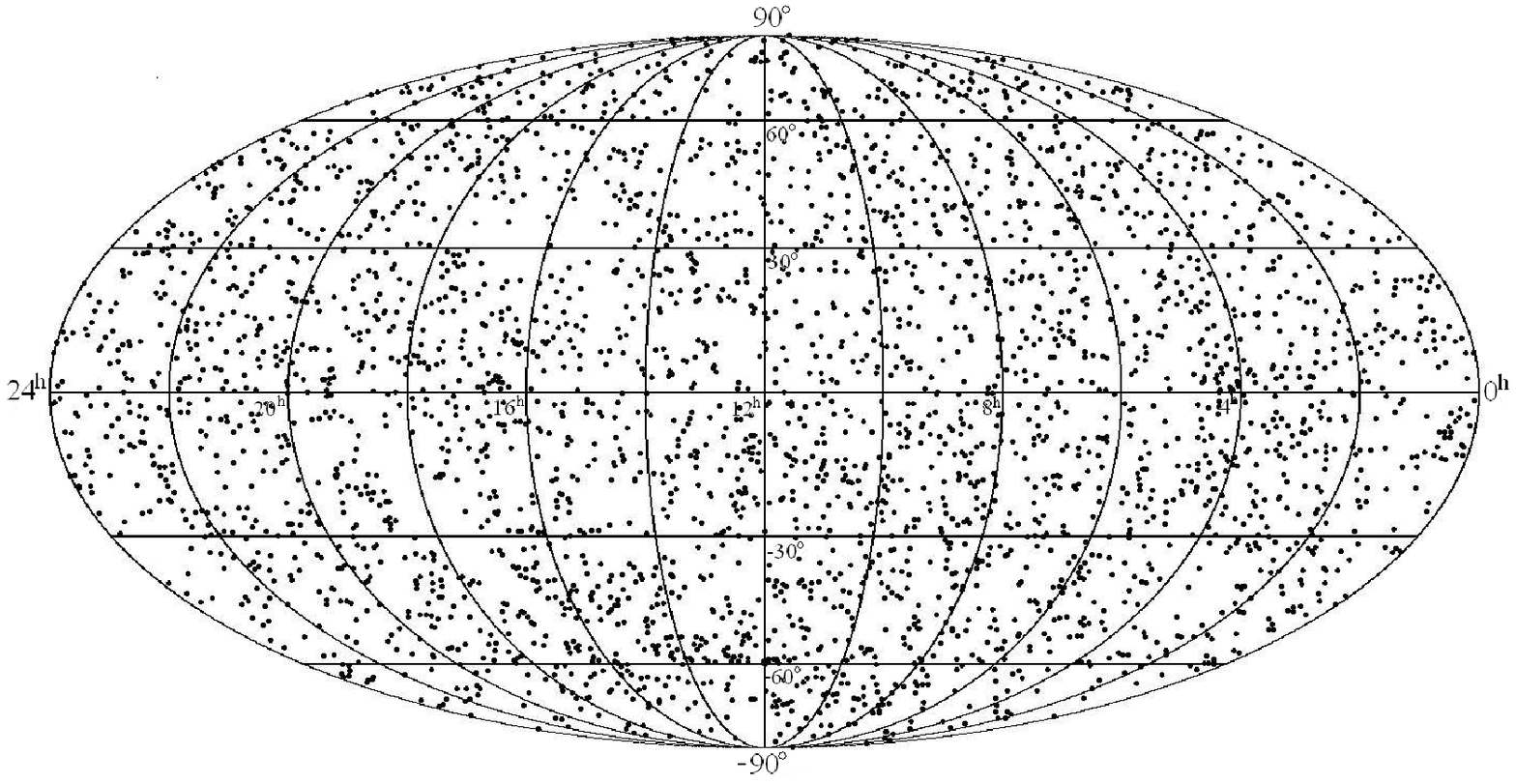}
\includegraphics[width=16cm, bb=16 538 507 805, clip]{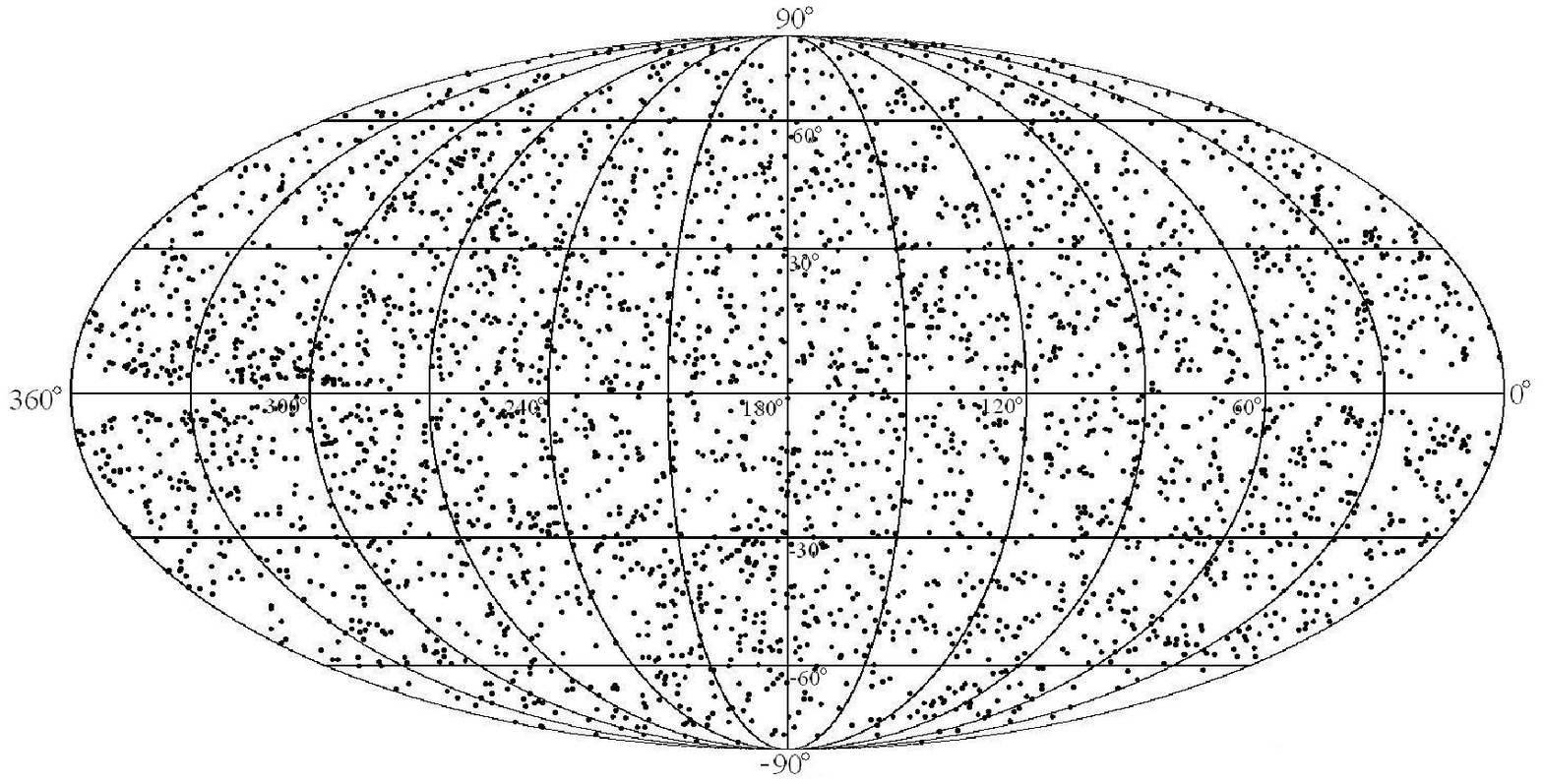}
\bigskip
\caption{Celestial distribution of 3227 isolated galaxies in
equatorial (the upper panel) and Galactic (the lower panel)
coordinates.}
\end{figure*}

\begin{figure*}[tbp]
\setcaptionmargin{5mm}
\onelinecaptionstrue
\includegraphics[width=14cm, bb=5 9 297 219, clip]{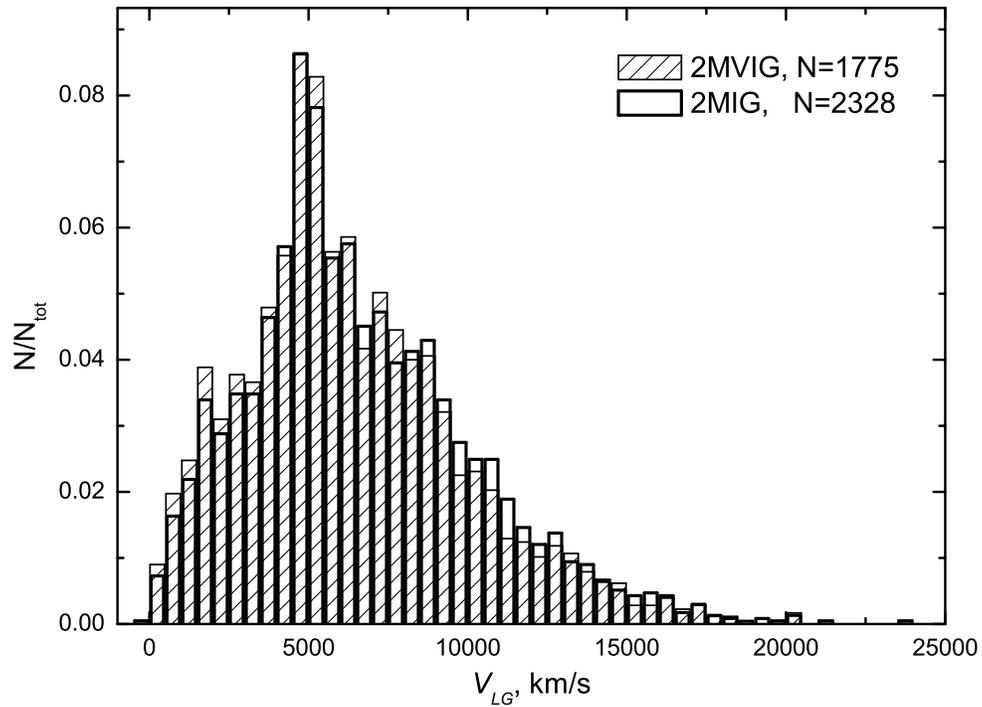}
\captionstyle{normal} \caption{Distribution of radial velocities
$V_{LG}$ of the 2MIG and 2MVIG galaxies.}
\end{figure*}

\begin{figure*}[tbp]
\setcaptionmargin{5mm}
\onelinecaptionsfalse
\includegraphics[width=13cm, bb=12 13 308 232, clip]{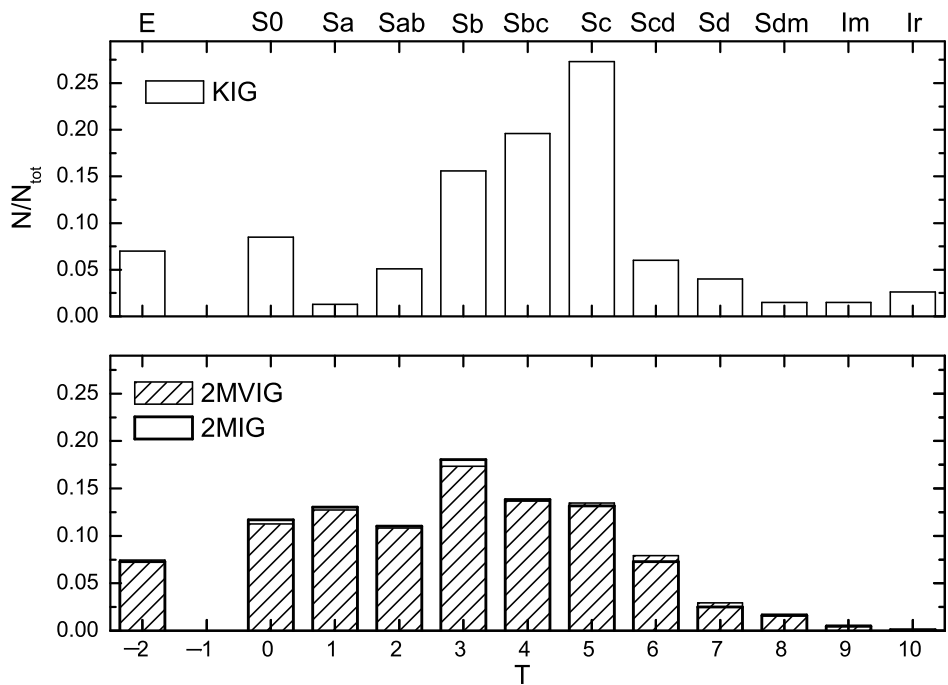}
\captionstyle{normal} \caption{Distribution of morphological types
of isolated galaxies: KIG galaxies
from~\cite{Hernandes-Toledo:Mitronova_n} and 2MIG galaxies.}
\end{figure*}

\begin{figure*}[tbp]
\setcaptionmargin{5mm}
\onelinecaptionsfalse
\includegraphics[width=13cm, bb=12 10 285 223, clip]{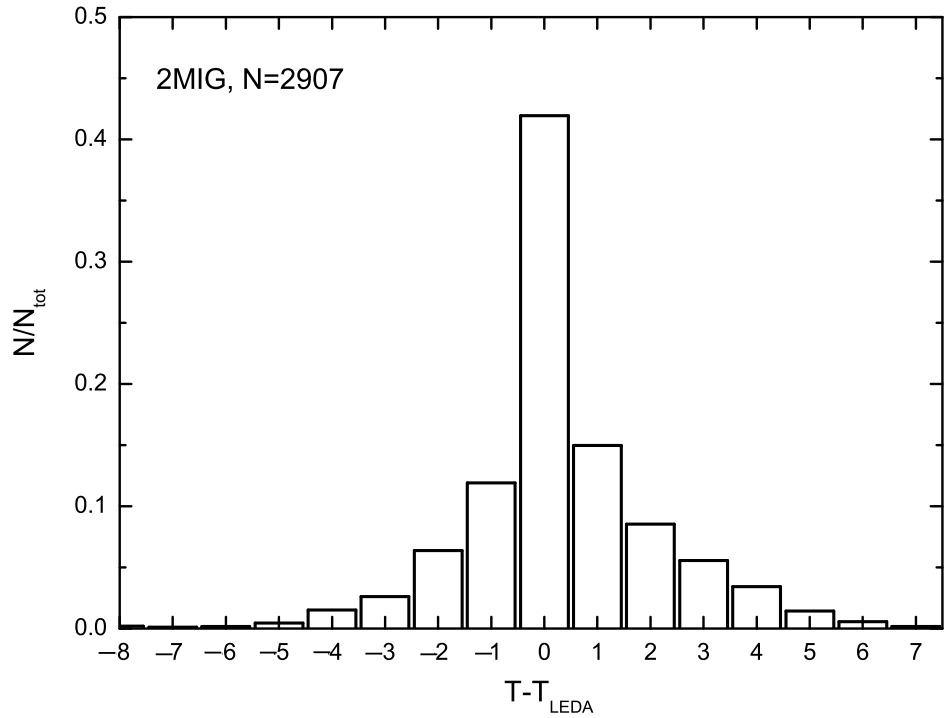}
\captionstyle{normal} \caption{Comparison of morphological type
estimates for  2MIG galaxies obtained in this paper with those
listed in the  LEDA database.}
\end{figure*}

\begin{figure*}[tbp]
\setcaptionmargin{5mm}
\onelinecaptionsfalse
\includegraphics[width=13cm, bb=12 10 285 226, clip]{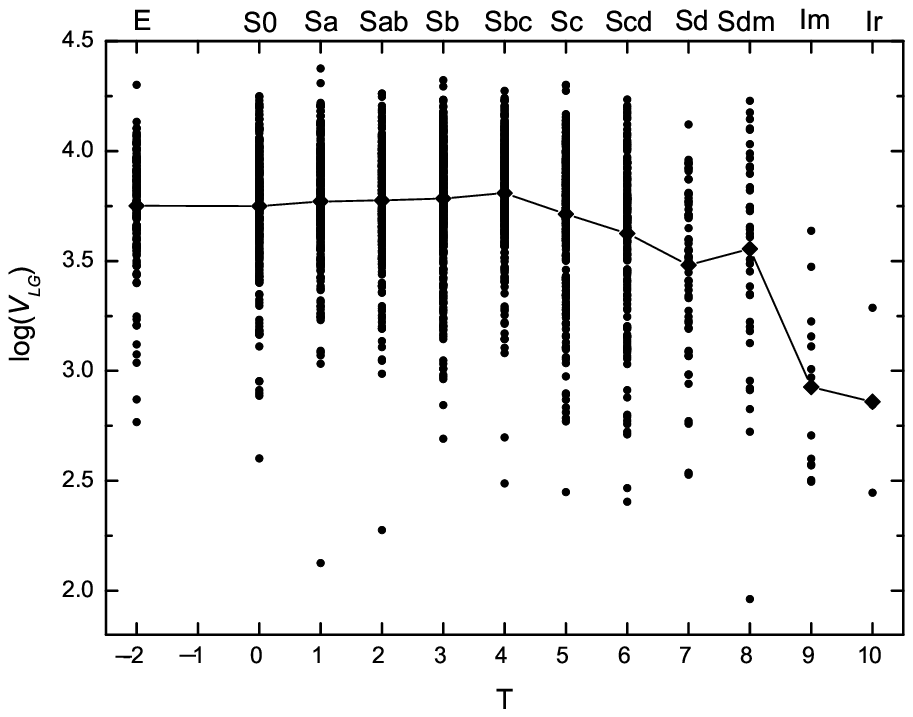}
\captionstyle{normal} \caption{The ``radial
velocity--morphological type'' relation. The line connects the
median radial velocity values for each type.}
\end{figure*}

Individual $K_{\rm s}$-magnitude differences (Fig.\,4) between
isolated galaxies and their nearest neighbors are distributed
asymmetrically in the  $[-5.5^m, +4.5^m]$ interval with a maximum
near $+2.0^m$. About 1/3 of all 2MIG galaxies have significant
members that are brighter than the isolated galaxies themselves.

Figure~5 demonstrates differential distributions of angular
diameters $a_K$ (in arcsec) of a number of isolated galaxies and
their most significant neighbors. The minimum angular diameter
condition imposed on the galaxies in the 2MASS XSC catalog
\mbox{$(a_K)_{min}= 10\arcsec$} combined with condition (2)
results in a formal requirement that only galaxies with diameters
$a_K> 40\arcsec$ can be viewed as candidate isolated galaxies. We
imposed a softer constraint \mbox{$a_K \geq 30\arcsec$.} It is
evident from the data shown in Fig.\,5 that the use of the  $a_K>
40\arcsec$ constraint would result in loss of 1230 galaxies, which
make up for 38\% of our sample.

In Fig.~6 one can see the distribution of the dimensionless
parameter $2s$ for the nearest significant neighbors of 3227 2MIG
galaxies. The mean and the standard deviation of this parameter
for the 2MIG galaxies are equal to \mbox{$\langle 2s\rangle=81.1$}
and $SD=21.3$, and the corresponding values for the 2MVIG galaxies
are equal to $81.9$ and $21.8$, respectively, i.e., differ
\mbox{insignificantly} for the two samples considered.

\newcommand{\2}{2MIG}

\section{BASIC PROPERTIES OF \2 SAMPLE}

Figure~7 represents the relation between the apparent magnitude
\mbox{$K_{\rm s}$} and the logarithm of angular
diameter~~$a_K$~for~3227~\2-galaxies. It can be \linebreak
characterized by a direct regression line \linebreak
\mbox{$\langle K_{\rm s} | a \rangle = -3.77\log a+17.29$} (the
solid line) with a standard deviation of $SD=0.49^m$; the inverse
regression $\langle \log a | K_{\rm s} \rangle = -0.17K_{\rm s}
+3.53$ is shown by the dotted line.

In Fig.~8 we demonstrate the cumulative distribution of apparent
magnitude $K_{\rm s}$ for different samples: for all 1.65\,
million 2MASS~XSC galaxies used in testing the isolation
conditions ({\it 1}); for 51572 galaxies brighter than~$K_{\rm s
}=12.0^m$ with angular diameters $a_K \geq 30\arcsec$, which were
tested for isolation ({\it 2}); for 3227 \2 galaxies with~$K_{\rm
s }<12.0^m$ and \mbox{$a \geq 30\arcsec$} included into the
catalog of isolated galaxies ({\it 3}); for the 2493 most isolated
galaxies of the~2MVIG sample ({\it 4}). The two parallel lines
correspond to the uniform distribution with a slope of~$0.6K_{\rm
s}$. It is evident from this figure that faint extended galaxies
of the 2MASS~survey follow the uniform distribution and their
excess at the bright end  \mbox{($K_{\rm s }<9^m$)} is due to the
presence of the Local Supercluster. A comparison of distributions
{\it 1} and {\it 2} leads us to conclude that imposing the
constraint~$a_K\ge30''$ on candidate isolated galaxies
with~$K_{\rm s } \leq 12^m\!.0$ reduces their number by
about~$40\%$. Isolated galaxies of the entire \2 catalog and
the~2MVIG subsample agree well with the uniform distribution and
demonstrate only a small deficit of isolated objects among bright
galaxies (as expected due to the effect of the Local
Supercluster). Thus, our criterion identifies about the same
fraction of isolated objects both among nearby and distant
galaxies. Note that the sample of isolated galaxies identified in
the~SDSS survey~[3] exhibits a strong dependence of a fraction of
isolated galaxies on their apparent magnitude, i.e., it is subject
to a significant selection effect.

Figure~9 shows the sky distribution of \2 galaxies in  equatorial
and Galactic coordinates. The distribution appears to be quite
uniform with no appreciable excess or deficit of galaxies in the
regions of the well-known Virgo, Fornax, or~Coma clusters. The
average number density of isolated galaxies is almost independent
on the Galactic latitude, thereby corroborating the efficiency of
the 2MASS~survey, which depends only slightly on the Galactic
absorption. However, the distribution of~\2 galaxies exhibits
voids and clumps in the direction towards the Galactic center (at
Galactic longitudes  $l=\pm30\degr$ and latitudes
\mbox{$b=\pm7\degr$)}, that are due to both interstellar
extinction and high density of stellar images, which sometimes
form false 2MASS~XSC sources mistaken for galaxies. For these
reasons the region in the vicinity of the Milky Way center
mentioned above should be excluded when performing a rigorous
statistical analysis of the \mbox{\2 catalog.}

More than 70\% of \2 galaxies have measured radial velocities. In
Fig.~10 one can find the distribution of radial velocities of
2328~isolated galaxies of our catalog with respect to the centroid
of the Local Group. The hatched area shows the distribution of
1775~galaxies of the~2MVIG subsample. The distribution peaks
at~5000\,km/s and the mean radial velocity is equal to~6570 and
6360\,km/s for the \2 and 2MVIG catalogs, respectively. Thus, the
characteristic depth of the new sample of isolated galaxies is
about the same as that of the KIG catalog \mbox{(6624\,km/s
\cite{Verley:Mitronova_n}).} Note that the mean radial velocity of
the nearest significant members of~\2 galaxies (12000\,km/s) is
almost twice as high as the velocities of the isolated galaxies,
i.e., most of the projected neighbors are background objects,
which are unassociated with isolated galaxies.

The lower panel in Fig.~11 presents the distribution of the
occurrence rates of \2 galaxies by morphological types. The
hatched histogram shows the distribution of more isolated 2MVIG
galaxies. The occurrence rates of different morphological types in
the ~\2 and~2MVIG samples differ insignificantly. Elliptical and
lenticular galaxies make up for about~19\% of the sample, whereas
the fraction of irregular galaxies ($T=9, 10$) does not
exceed~1\%. According to initial
estimates~\cite{Kar+Kara:Mitronova_n}, the fraction of~E and~S0
galaxies in the KIG is equal to 18\%, which practically coincides
with the fraction of these galaxies in the~\2 catalog. The
fraction of irregular galaxies in the KIG is equal to  10\%, which
is one order of magnitude higher than in the new catalog.
Recently, Sulentic et al.~\cite{Sulentic:Mitronova_n} and
Hernandez-Toledo et al.~\cite{Hernandes-Toledo:Mitronova_n}
revised the morphological types of KIG galaxies using DSS-2 and
SDSS images. According to the data of Hernandez-Toledo et
al.~\cite{Hernandes-Toledo:Mitronova_n}, that we demonstrate in
the upper panel of Fig.\,11, E and~S0-type galaxies make up for
16\%, and irregular galaxies for 4\% of all the KIG galaxies. It
follows from the comparison of the upper and lower panels in
Fig.\,11 that the fraction of elliptical and lenticular galaxies
in the ~\2 catalog is about the same as in the KIG, whereas the
occurrence rate of early-type \linebreak spirals (Sa, Sab) in
the~\2 catalog is appreciably higher than in the KIG. These
differences appear quite expected given the low sensitivity of the
infrared survey to blue extended structures of low surface
brightness, which are typical of late-type galaxies. To assess the
accuracy of the galaxy morphological type findings, we compared
our estimates with the morphological types of 2907~\2 galaxies
given in the~LEDA database. Figure~12 demonstrates the
distribution of differences between the two independent estimates.
In~69\% of the cases, the estimates differ by no more than one
subtype. However, large discrepancies occur for some galaxies.
Significant differences between morphological type estimates were
found, e.g., in cases where a regularly shaped blue compact galaxy
was mistaken for an elliptical galaxy. The availability of
infrared images for all~\2-galaxies allowed us to distinguish BCG
and E-galaxies with more confidence.

The above occurrence rates of different galaxy types in both~\2
and ~KIG refer to catalog samples. The corresponding frequencies
for a fixed volume may differ from the above values since galaxies
of different types have different luminosities. Figure~13
demonstrates the relation between the radial velocities
of~\2-galaxies and their morphological type. It is evident from
this figure that the median radial velocity remains the
same---about $5600\,$km/s for the galaxies of types spanning
from~E to~Sc, and decreases towards later types: the median radial
velocity for Im and Ir galaxies is as low as $800\,$km/s. This
radial velocity trend reflects the variation of average infrared
luminosity along the morphological type sequence. The volume
occupied by the catalogued irregular galaxies is about 300~times
smaller than the volume occupied by~E--Sc-type galaxies and
therefore the occurrence frequency of the former in the unit
volume must be much higher than their occurrence frequency in the
catalog. The data from the new catalog of isolated galaxies in the
Local Supercluster~\cite{KarachentsevEtAl:Mitronova_n} corroborate
this evident conclusion. The fraction of irregular galaxies in
this radial velocity limited sample of galaxies amounts to $30\%$.

\section{CONCLUDING REMARKS}

We undertook an automatic search for isolated galaxies among the extended sources of the  Two Micron
All Sky Survey (2MASS XSC) with  $K_{\rm s}$ magnitudes in the interval $4.0^m< K_{\rm s} \leq 12.0^m$
and infrared angular diameters  $a_K \geq 30\arcsec$.

To test the isolation of galaxies, we used more than one and a
half million  2MASS~XSC objects with apparent magnitudes~$K_{\rm s
}<14^m\!.5$. From the resulting sample of 4045~candidate isolated
galaxies we excluded the objects that proved to be planetary
nebulae or star clusters ($N=250$). Since the 2MASS~survey is
insensitive to blue low surface brightness galaxies, we inspected
the  neighborhood of each candidate galaxy on the  \mbox{POSS-I,}
POSS-II, and~SDSS digital sky surveys. As a result, we found
2493~very isolated galaxies (the 2MVIG sample) that have no
significant companions (according to Karachentseva's criterion)
neither in the infrared nor at optical wavelengths. The use of the
available radial velocity data revealed a total of 567~galaxies
that are not isolated, since they have significant neighbors or
belong to groups. The remaining galaxies without radial velocity
estimates ($N=734$) along with the~2MVIG sample made up the
combined catalog of isolated galaxies (\2) containing a total of
~\mbox{$N=3227$} objects.

The celestial distribution of \2 galaxies appears to be quite
uniform without appreciable excess or deficit of galaxies in the
regions of nearby clusters and groups. Isolated galaxies of our
catalog account for 6\% objects among the galaxies brighter
than~$K_{\rm s }=12^m.0$ with diameters~$a_K \geq 30\arcsec$, and
the fraction of~\2 galaxies is about the same both among bright
and faint (distant) galaxies. The mean radial velocity of isolated
galaxies is equal to~6500\,km/s, making the~\2 sample comparable
in depth with the Catalog of isolated
galaxies~(KIG)~\cite{Kara:Mitronova_n}. The fraction of isolated~E
and~S0 galaxies in the new catalog (19\%) is almost the same as in
the KIG, whereas the fraction of late-type spirals and irregulars
in the~\2 catalog is much less than in the KIG.

The catalog reported in this paper can be viewed as a homogeneous
reference sample for further studies of the effects the
environment imposes on the structure and evolution of galaxies.

\newpage
\footnotesize \setcaptionmargin{10mm} \onelinecaptionsfalse
\captionstyle{nonumber} \tiny
\begin{longtable*}{c|c|c|c|c|c|c|c|c|c}
\caption{{\bf Table.}~The 2MIG catalog of isolated galaxies}\\
 \hline 
2MIG    &       RA, DEC (J2000)   & Name &       $r$       &       $K_{\rm s}$      &      $2s$      &       $V_h$      &       T       &  N & Comments  \\
\hline
(1)& (2)& (3)&(4)&(5)&(6)&(7)&(8)&(9)&(10)\\
\hline
\endfirsthead
\caption{{\bf Table.}~(Contd.)}\\

\hline
2MIG    &       RA, DEC (J2000)   & Name       &       $r$       &  $K_{\rm s}$  &      $2s$      &       $V_h$      &       T       &  N & Comments   \\
\hline
(1)& (2)& (3)&(4)&(5)&(6)&(7)&(8)&(9)&(10)\\
\hline
\endhead

\hline
\endfoot

\hline 
\endlastfoot
1    & 00002508+0751138 & UGC12892        &  23.7  & 11.12 &   66  &        &  2  &   &  Bar, ring     \\
2    & 00005858--3336429 & ESO349--017      &  21.6  & 11.55 &   61  &  6909  &  5  &   &  Vc        \\
3    & 00015230+4020109 & UGC12917        &  20.9  & 11.62 &   95  &        &  3  & 2 &  Bar, ring        \\
4    & 00020314--4521288 & PGC130018       &  17.9  & 11.51 &   76  &  11639 &  3  &   &          \\
5    & 00030565--0154495 & UGC00005        &  30.1  & 10.34 &   99  &  7296  &  4  & 1 &  Bar, HII, KIG1, IRAS \\
6    & 00034871--4337058 & PGC262          &  18.8  & 11.73 &   74  &  9076  &  2  &   &  AM0001-435      \\
7    & 00041078--1313190 & PGC941042       &  17.2  & 11.89 &   63  &        &  3  & 2 &          \\
8    & 00050536--0705363 & IC1528          &  41.4  & 10.39 &   94  &  3768  &  3  & 2 &  HII     \\
9    & 00051322--1130093 & IC1529          &  21.9  & 10.37 &  113  &  6751  &  0  & 2 &  Pec, ring        \\
10   & 00054271--7542251 & ESO028--009      &  26.6  & 11.07 &  106  &  6042  &  4  & 1 &  IRAS  \\
11   & 00081466+0746487 & UGC00067        &  32.1  & 10.40 &   78  &  11833 &  2  &   &  Ring    \\
12   & 00083428--1056579 & MCG--02--01--028   &  20.5  & 11.84 &  101  &  9109  &  2  & 1 &  Bar, ring        \\
13   & 00083453--3351299 & NGC0010         &  56    & 9.48  &   67  &  6806  &  3  &   &  Bar, IRAS, Vc      \\
14   & 00084249+3726523 & NGC0011         &  43    & 9.98  &  181  &  4390  &  1  & 2 &  IRAS  \\
15   & 00085471+2349009 & NGC0009         &  22.2  & 11.80 &  125  &  4527  &  3  &   &  Pec, HII, KIG6, IRAS    \\
16   & 00090246+2137279 & NGC0015         &  26.9  & 10.42 &   64  &  6330  &  1  & 1 &          \\
17   & 00090345--3254509 & ESO349--033      &  26    & 10.99 &   61  &  6892  &  3  & 2 &  IRAS  \\
18   & 00090421+1055081 & UGC00081        &  29.3  & 10.72 &   76  &  6674  &  3  & 3 &          \\
19   & 00095654--2457472 & NGC0024         &  83    & 9.22  &  147  &  554   &  5  &   &  IRAS  \\
20   & 00101611+3206223 & CGCG499--061     &  15.3  & 11.72 &   71  &  19372 &  3  & 1 &          \\
21   & 00110081--1249206 & IC0002          &  20.3  & 11.24 &   80  &        &  2  &   &  IRAS  \\
22   & 00110634+0240406 & CGCG382--030     &  21.4  & 11.58 &   72  &  12760 &  4  &   &  Ring, KIG7       \\
23   & 00140398--2310555 & NGC0045         &  49.3  & 10.07 &   79  &  466   &  8  &   &  Bar, IRAS      \\
24   & 00141284+2245591 & CGCG478--044     &  16.4  & 11.02 &   80  &        &  1  & 1 &          \\
25   & 00144279--6019425 & NGC0053         &  38.8  & 10.30 &  112  &  4572  &  2  &   &  Bar, ring        \\
26   & 00145057--8659351 & ESO002--006      &  31.7  & 9.92  &  167  &        &  --2 &   &          \\
27   & 00151647--5714412 & ESO111--022      &  22    & 11.13 &   72  &  9800  &  3  &   &  Ring, Vc    \\
28   & 00161479+1019565 & UGC00151        &  23.9  & 10.36 &   84  &  5254  &  --2 &   &  KIG13   \\
29   & 00165087--0516060 & MCG--01--01--064   &  32.7  & 10.64 &   81  &  3943  &  1  & 1 &  Bar, LINER, HII, IRAS    \\
30   & 00170507+4209410 & UGC00158        &  19.1  & 11.69 &   65  &  5065  &  3  &   &          \\
31   & 00170970--0342489 & MCG--01--01--067   &  19.8  & 10.83 &   67  &  10959 &  --2 &   &          \\
32   & 00175470--4755408 & ESO194--001      &  20.6  & 11.87 &   67  &  11450 &  5  & 1 &       \\
33   & 00181211+1311321 & MCG+02--01--031   &  24.2  & 10.99 &   92  &  4131  &  1  &   &  IRAS \\
34   & 00182395+4843543 & UGC00171        &  21.5  & 10.54 &   62  &  5266  &  4  &   &  Pec, HII, IRAS  \\
35   & 00194400--1407184 & IC0009          &  15.3  & 11.36 &   87  &  12622 &  3  & 1 &  Ring, Sy2, HII, IRAS     \\
36   & 00194874+2346214 & IC1540          &  28.3  & 10.85 &   65  &  5827  &  3  &   &  Bar     \\
37   & 00200006--0620013 & MCG--01--02--001   &  24.2  & 11.38 &   73  &  3709  &  2  &   &  Bar, pec \\
38   & 00214374--6142399 & ESO112--001      &  29.2  & 11.16 &   61  &        &  5  &   &          \\
39   & 00215111--0929321 & MCG--02--02--005   &  20.8  & 11.37 &  113  &  6267  &  3  & 1 &  Ring, IRAS     \\
40   & 00220122+4908003 & CGCG549--038     &  23.5  & 10.95 &   74  &  5144  &  4  &   &  Pec, HII, IRAS  \\
41   & 00223386--0829109 & MCG--02--02--007   &  21.9  & 11.07 &  110  &  5692  &  4  & 1 &          \\
42   & 00231109--5937029 & ESO111--026      &  17    & 11.36 &  102  &        &  2  & 1 &          \\
43   & 00233603+2051113 & UGC00225        &  17    & 11.28 &   78  &        &  1  & 1 &          \\
44   & 00235461--3232103 & NGC0101         &  32.9  & 10.77 &  122  &  3383  &  5  &   &  Bar, HII, IRAS  \\
45   & 00243651--1357229 & NGC0102         &  28.8  & 10.74 &   63  &  7332  &  0  &   &  Bar, ring        \\
46   & 00245879+4339459 & UGC00236        &  17.4  & 10.76 &   94  &  5104  &  --2 &   &  KIG20   \\
47   & 00250335+3120411 & UGC00238        &  33.9  & 10.57 &   78  &  6766  &  4  & 1 &  LINER, IRAS  \\
48   & 00252991+4555181 & UGC00243        &  57.8  & 9.61  &   78  &  5076  &  3  &   &  IRAS, Vc  \\
49   & 00260321--0720047 & PGC172039       &  15.1  & 11.29 &   66  &  15888 &  0  & 1 &        \\
50   & 00261744--0429323 & MCG--01--02--014   &  31.2  & 10.98 &   92  &  3983  &  2  &   &  Bar, pec, HII, IRAS  \\
51   & 00262976--6003220 & PGC127809       &  21.4  & 11.33 &   83  &  4749  &  3  & 1 &  AM0024-602, IRAS       \\
52   & 00265513--4438186 & ESO242--009      &  16.1  & 11.80 &   64  &        &  9  & 2 &          \\
53   & 00265761--5658408 & NGC0119         &  27.3  & 10.00 &   71  &  7430  &  --2 &   &          \\
54   & 00272276+1050542 & UGC00266        &  18.5  & 11.33 &   61  &        &  1  & 2 &          \\
55   & 00281783--0929343 & MCG--02--02--020   &  18.8  & 11.91 &   61  &  16959 &  3  &   &          \\
56   & 00290063--0819062 & MCG--02--02--025   &  15.2  & 11.24 &   68  &  6110  &  0  &   &          \\
57   & 00291681+5319125 & PGC2437721      &  18.5  & 11.05 &   82  &        &  1  &   &          \\
58   & 00294166--5131145 & ESO194--021      &  35.9  & 9.12  &  123  &  3496  &  --2 &   &          \\
59   & 00294368+2128365 & IC1552          &  27.9  & 10.99 &  100  &  5600  &  5  &   &  KIG23, IRAS    \\
60   & 00300543--6013492 & PGC143541       &  18.8  & 11.36 &   81  &  11923 &  1  &   &          \\
61   & 00302865+0551405 & CGCG409--021     &  19.4  & 11.10 &   64  &  7087  &  0  &   &          \\
62   & 00304038--2842454 & ESO410--015      &  22.6  & 10.99 &   64  &  7307  &  1  &   &  Vc      \\
63   & 00304382--5900258 & ESO112--006      &  15.5  & 11.44 &   72  &  8642  &  2  & 1 &  IRAS  \\
64   & 00313584+1436442 & UGC00316        &  26    & 11.51 &   66  &  11432 &  6  &   &  Vc    \\
65   & 00314682+6817323 & PGC137056       &  16.4  & 11.69 &   62  &        &  5  &   &        \\
66   & 00324265--1119054 & MCG--02--02--049   &  44.6  & 10.35 &   62  &  4031  &  5  &   &  IRAS  \\
67   & 00331028--1308462 & MCG--02--02--051   &  15.8  & 11.81 &   64  &  6179  &  4  &   &  Bar, ring, IRAS \\
68   & 00333080+2254100 & CGCG479--039     &  15.8  & 11.48 &   98  &  4599  &  0  &   &  KIG27, IRAS    \\
69   & 00342461+1216066 & IC0031          &  32.4  & 10.37 &   79  &  9515  &  2  & 1 &          \\
70   & 00344675--0823473 & NGC0157         &  95.5  & 7.68  &   65  &  1673  &  4  &   &  Bar, IRAS      \\
71   & 00345798--5133233 & PGC129232       &  20.6  & 11.85 &   71  &        &  4  & 2 &          \\
72   & 00360908+5522418 & PGC137012       &  21.2  & 11.35 &   64  &        &  3  &   &          \\
73   & 00372152--1956032 & NGC0171         &  64.5  & 9.39  &   62  &  3884  &  2  &   &  Bar, ring, VV791A, IRAS  \\
74   & 00373987+1021287 & IC0035          &  20    & 11.13 &   95  &  4587  &  6  & 2 &  KIG30   \\
75   & 00375565+0454408 & CGCG409-049     &  26.4  & 11.25 &   75  &  8489  &  5  &   &  IRAS, Vc \\
76   & 00375769+0838068 & NGC0180         &  55.2  & 9.73  &   69  &  5279  &  4  & 1 &  Bar, IRAS      \\
77   & 00382373+1502223 & UGC00386        &  16.3  & 11.18 &   64  &  5376  &  1  &   &  MRK343, IRAS, Vc   \\
78   & 00383973+1724113 & UGC00393        &  19.2  & 11.38 &   91  &  5432  &  3  &   &  Vc            \\
79   & 00391554--4304315 & ESO242--023      &  34.3  & 10.72 &   67  &  4026  &  5  &   &  Pec, IRAS      \\
80   & 00393031+2304220 & CGCG479--053     &  16.6  & 11.55 &   60  &        &  2  & 1 &          \\
81   & 00393535--4717285 & 2MFGC472        &  15.7  & 11.86 &   82  &        &  3  & 1 &          \\
82   & 00403145--4559074 & ESO242--024      &  31.3  & 10.39 &   99  &  3695  &  1  &   &  Bar, IRAS      \\
83   & 00410364+3143576 & UGC00433        &  50.8  & 10.22 &   89  &  4654  &  5  &   &  IRAS  \\
84   & 00411934+6855542 & PGC137148       &  15.6  & 11.80 &   98  &        &  3  & 1 &          \\
85   & 00432575--5010580 & NGC0238         &  53.1  & 10.06 &   96  &  8614  &  3  &   &  Bar, ring, IRAS, Vc \\
86   & 00433238+1420334 & NGC0234         &  34.4  & 9.62  &  151  &  4457  &  5  & 1 &  IRAS  \\
87   & 00441293--1235316 & MCG--02--03--004   &  20.3  & 11.15 &   84  &  6784  &  0  & 1 &  IRAS  \\
88   & 00450202--6045373 & ESO112--009      &  18.7  & 10.92 &   83  &  10500 &  0  & 1 &          \\
89   & 00454643--1535487 & NGC0244         &  18    & 11.32 &   81  &  941   &  9  &   &  VV728   \\
90   & 00471276+2908110 & 2MFGC00567      &  16.5  & 11.34 &   71  &  5706  &  2  &   &  IRAS, Vc  \\
91   & 00475430+6807433 & 2MFGC574        &  37    & 10.86 &   96  &        &  5  &   &          \\
92   & 00480150+0817494 & NGC0257         &  40.3  & 9.66  &   78  &  5276  &  4  & 1 &  IRAS  \\
93   & 00484212--4040202 & 2MFGC00586      &  20.1  & 11.96 &   97  &  10038 &  5  &   &          \\
94   & 00484744--4609041 & PGC130104       &  15.4  & 11.60 &   61  &  15860 &  0  & 1 &          \\
95   & 00493452--4652279 & ESO243--002      &  20.2  & 11.62 &  112  &  8886  &  5  &   &          \\
96   & 00493887+2255564 & UGC00506        &  24.6  & 10.33 &   79  &  7462  &  --2 &   &          \\
97   & 00494975--3520031 & ESO351--010      &  18.8  & 11.63 &   92  &        &  4  & 1 &          \\
98   & 00500923--1326404 & PGC937908       &  16.9  & 11.67 &   93  &        &  1  & 2 &          \\
99   & 00500956--0511376 & NGC0268         &  32    & 10.58 &   87  &  5477  &  4  &   &  Bar, IRAS    \\
100  & 00501956+6702517 & PGC2796450      &  22.3  & 10.64 &   92  &        &  0  &   &          \\
101  & 00502252+1142376 & UGC00513        &  16.3  & 11.86 &   66  &  11901 &  3  & 1 &          \\
102  & 00513111--3739192 & MCG--06--03--001   &  16.8  & 11.10 &   75  &  7005  &  0  & 1 &          \\
103  & 00521377+4419514 & UGC00530        &  25.4  & 10.74 &   99  &  5331  &  1  & 1 &          \\
104  & 00533211--5806256 & IC1597          &  21    & 11.93 &   67  &  5053  &  3  & 2 &  Bar, pec, ring    \\
105  & 00543974--6317016 & ESO079--010      &  21.5  & 11.30 &   67  &  5662  &  5  & 2 &  IRAS  \\
106  & 00544282+2131215 & IC1596          &  18.9  & 11.63 &   90  &  2675  &  3  & 1 &  KIG39   \\
\end{longtable*}
\twocolumngrid
\normalsize

\begin{acknowledgments}
We are grateful to D.~I.~Makarov for useful discussions and
assistance with the Pleinpot environment package. This research
makes use of the  DSS ({\tt http://archive.eso.org/dss/dss}) and
SDSS \linebreak ({\tt http://www.sdss.org}) digitized sky surveys
and the  HYPERLEDA ({\tt http://leda.univ-lyon1.fr/}) and NED
({\tt http://nedwww.ipac.caltech.edu/}) databases. Olga Melnyk
acknowledges the financial support from the Belgian Science
Policy. This work was supported in part by the Russian Foundation
for Basic Research (projects \mbox{nos\,06-02-04017-NNIO-a,}
07--02--00005-a, and~09--02--90414-Ukr-a), and by the State
Foundation for Basic Research of the Ministry of Science and
Education of Ukraine (project no. F28.2.059).
\end{acknowledgments}

\end{document}